\documentstyle[12pt,epsfig]{article}

\begin{document}

\begin{titlepage}

\begin{flushright}
{\sc TRI-PP-95-06}\\
{\sc hep-ph/9503285}\\[.2in]
{\sc June 19, 1995}\\ [.2in]
{\sc Revised Version}\\[.5in]
\end{flushright}

\begin{center}
{\LARGE
The contribution of the $J/\psi$ resonance
to the radiative $B$ decays}\\ [.5in]
{\large
Jo\~{a}o M. Soares}\\ [.1in]
{\small
TRIUMF, 4004 Wesbrook Mall, Vancouver, BC Canada V6T 2A3}\\ [.5in]

{\normalsize\bf
Abstract}\\ [.2in]
\end{center}

{\small
The radiative decays of the $B$ mesons may have a significant contribution
from the transition $b \rightarrow s J/\psi$ followed by the $J/\psi$-photon
conversion. The size of this contribution is re-analysed in the light of a
phenomenological model for the weak $bsJ/\psi$ vertex, and a modified
$J/\psi$-photon interaction that is manifestly gauge invariant. Predictions
for both inclusive and exclusive cases are obtained, but large uncertainties
still remain.
}

\end{titlepage}

\section{Introduction}

The CLEO Collaboration has observed the exclusive radiative decays of charged
and neutral $B$-mesons into $K^\ast$ \cite{CLEO:exclusive}, with an average
branching ratio
\begin{equation}
BR(B \rightarrow K^\ast \gamma) = (4.5 \pm 1.5 \pm 0.9) \times 10^{-5} .
\label{eq:1}
\end{equation}
More recently, the same experiment reported the first signs of the inclusive
decay $B \rightarrow \gamma + X_s$ \cite{CLEO:inclusive}, with the branching
ratio
\begin{equation}
BR(B \rightarrow \gamma + X_s) = (2.32 \pm 0.57 \pm 0.35)
\times 10^{-4} .
\label{eq:2}
\end{equation}
At the origin of these decays is predominantly the spectator process involving
the $bs\gamma$ vertex. In the Standard Model, the short distance contribution
to the vertex occurs at the 1-loop level, but it is sizeable due to the large
top-quark mass and an important QCD enhancement \cite{qcd:bsgamma}. It can be
calculated perturbatively, and the QCD corrections have been included in the
leading logarithm approximation \cite{lla:bsgamma}. The uncertainty in this
result is mostly due to the choice of the scale at which to calculate the QCD
corrections; with the full next-to-leading order calculation completed, this
error should be substantially smaller \cite{nlo:bsgamma}. However, it is
possible that a significant long distance contribution to the $bs\gamma$
vertex exists, due to the process $b \rightarrow s J/\psi \rightarrow s
\gamma$. The weak decay of the $b$-quark that produces the $J/\psi$ meson
occurs at tree level; the $J/\psi$ in turn couples to the photon, as in the
$J/\psi \rightarrow e^+ e^-$ decay mode. For the inclusive decay, a naive
estimate gives
\begin{equation}
|A(b \stackrel{J/\psi}{\rightarrow} s \gamma)| \sim
|A(b \rightarrow s J/\psi)|\: e g_{J/\psi\gamma}\: \frac{1}{m_{J/\psi}^2}\: ,
\label{eq:3}
\end{equation}
for the $J/\psi$ contribution to the decay amplitude. The strength of the
$J/\psi$-photon conversion, $g_{J/\psi\gamma} = 0.82$ ${\rm GeV}^2$, is
measured from the rate for $J/\psi \rightarrow e^+ e^-$. Eq.\ \ref{eq:3} gives
a long distance contribution that is about $20\%$ of the observed $b
\rightarrow s \gamma$ amplitude. The analogous estimate for the exclusive
decay $B \rightarrow K^\ast \gamma$ gives a $J/\psi$ contribution in the same
proportion. This effect was first pointed out by Golowich and Pakvasa
\cite{GP:88} as the dominant long distance contribution to the radiative $B$
decays; a phenomenological model for the exclusive process $B \stackrel{J/\psi}
{\rightarrow} K^\ast \gamma$ was proposed in ref.\ \cite{GP:88}, and later
expanded in ref.\ \cite{GP:94} (the analogous effect in the $B \rightarrow
\rho \gamma$ decay was discussed recently by Cheng \cite{Cheng:94}). The
inclusive case was considered by Deshpande, Trampetic and Panose in ref.\
\cite{DTP:88}, where a model could not be found that would satisfy gauge
invariance and give a non-zero result. More recently, one such model was
suggested by Deshpande, He and Trampetic \cite{Desh:94}.

In this work, the mechanism behind the long distance contribution of the
$J/\psi$ to the $B$-meson radiative decays is re-analyzed, within a new
phenomenological approach. The analysis will be based on an effective
$bsJ/\psi$ vertex (section 2.1), parametrized by form factors that are to be
determined empirically, and a $J/\psi$-photon interaction (section 2.2),
modeled after the vector meson dominance (VMD) ideas \cite{VMD:Sakurai}. From
this description one derives both the amplitude for the inclusive process $b
\rightarrow s J/\psi \rightarrow s \gamma$, and that for the exclusive process
$B \rightarrow K^\ast J/\psi \rightarrow K^\ast \gamma$ (section 2.3). These
amplitudes are automatically gauge invariant, and vanish when the $bsJ/\psi$
vertex is calculated in the factorization approximation. Quantitative
predictions are derived (section 3), but significant uncertainties still
remain. This work was inspired in the recent analysis of ref.\ \cite{GP:94},
which also adopts a phenomenological approach to determine the size of the
long distance effect, for the case of the exclusive decay. The model and the
results obtained in here are however substantially different. The same is true
with respect to the other analyses that have appeared in the literature
\cite{Other:94}.

\section{The $J/\psi$ contribution to the radiative $B$ decays}

\subsection{The $bsJ/\psi$ vertex}

The amplitude for the inclusive decay $b \rightarrow s J/\psi$ is given by
\begin{equation}
A(b \rightarrow s J/\psi) = - < s J/\psi |H_{eff}| b > ,
\label{eq:4}
\end{equation}
where $H_{eff}$ is the effective Hamiltonian that describes the weak process
$b \rightarrow s c \overline{c}$:
\begin{equation}
H_{eff} = \frac{G_{F}}{\sqrt{2}} \: V_{cb} V_{cs}^\ast \:
( C_1 \: \overline{c}_\alpha \gamma_{\mu} L  c_\beta \:
\overline{s}_\beta \gamma^{\mu} L  b_\alpha +
C_2 \: \overline{c}_\alpha \gamma_{\mu}  L c_\alpha \:
\overline{s}_\beta \gamma^{\mu} L  b_\beta )
\label{eq:5}
\end{equation}
($L,R \equiv 1 \mp \gamma_5$). The Wilson coefficients $C_1$ and $C_2$ contain
the short distance QCD corrections. In the leading logarithm approximation,
for $\Lambda_{\overline{MS}}^{(5)} = 200$ MeV \cite{Pdg:94}, and at the scale
$\mu = 5.0$ GeV, they are \cite{SD:QCD}
\begin{equation}
C_1 = 1.117 \hspace{.3in} C_2 = - 0.266  .
\label{eq:6}
\end{equation}
The soft QCD effects in the hadronization of the $c$-$\overline{c}$ pair are
described in terms of form factors that parametrize the matrix element of
$H_{eff}$, in eq.\ \ref{eq:4}. For example, in the factorization prescription
\cite{fact:bspsi},
\begin{eqnarray}
< s J/\psi | \overline{c}_\alpha \gamma_{\mu} L c_\alpha \:
\overline{s}_\beta \gamma^{\mu} L b_\beta | b > &=&
3 < s J/\psi |\overline{c}_\alpha \gamma_{\mu} L c_\beta \:
\overline{s}_\beta \gamma^{\mu} L b_\alpha | b > \nonumber\\
&=& m_{J/\psi} f_{J/\psi} \varepsilon_\mu^\ast
\overline{u}_s \gamma^\mu L u_b  .
\label{eq:7}
\end{eqnarray}
The $J/\psi$ decay constant, $f_{J/\psi}$, is defined by $<0| \overline{c}
\gamma_\mu c |J/\psi> = m_{J/\psi} f_{J/\psi} \varepsilon_\mu$, and it is the
only form factor that enters the $b \rightarrow s J/\psi$ decay amplitude,
within factorization. From $\Gamma(J/\psi \rightarrow e^+ e^-) = (5.26 \pm
0.37)$ keV \cite{Pdg:94}, it follows that $f_{J/\psi} = 395$ MeV.

In all generality, however, one can write an effective $bsJ/\psi$ vertex that,
for on-shell quarks and with $m_s = 0$, is given by
\begin{eqnarray}
\Lambda^\mu_{bsJ/\psi} &=& - \frac{G_F}{\sqrt{2}} \: V_{cb}V_{cs}^\ast \:
(C_2 + \frac{1}{3} C_1)\:
[ g_0(k^2) \: k^\mu \not{k} L  \nonumber\\
& & + g_1(k^2) \: (k^2 g^{\mu\nu} - k^\mu k^\nu) \gamma_\nu L +
g_2(k^2) \: m_b i \sigma^{\mu\nu} k_\nu R ] ,
\label{eq:8}
\end{eqnarray}
where $k$ is the $J/\psi$ four-momentum. The motivation to adopt this more
general approach is the fact that the factorization result, $g_1(m_{J/\psi}^2)
= g_0(m_{J/\psi}^2)= f_{J/\psi}/m_{J/\psi}$ and $g_2(m_{J/\psi}^2) = 0$, gives
a very poor agreement with the data, for both the inclusive and the exclusive
decays \cite{fact:bspsi}. Indeed, at present, there is no satisfactory
theoretical description of the weak $b$ decay that produces the $J/\psi$
meson. In here, the form factors $g_1$ and $g_2$, at $k^2 = m_{J/\psi}^2$, are
to be determined empirically, from the data for the $B$-meson decays into
$J/\psi$. The term proportional to the form factor $g_0$ does not contribute
to the decay amplitudes, and so $g_0(m_{J/\psi}^2)$ will be left undetermined.
Notice that, unless this form factor vanishes, the $J/\psi$ meson couples to a
current
\begin{equation}
J^\mu = - \overline{s} \Lambda^\mu_{bsJ/\psi} b
\label{eq:9}
\end{equation}
that is not conserved.

\subsection{The $J/\psi$ contribution to the $bs\gamma$ vertex.}

The effective $bs\gamma$ vertex, for on-shell quarks and with $m_s = 0$, is
analogous to that in eq.\ \ref{eq:8}, but with an additional constraint from
gauge invariance. The contribution from the $c$-$\overline{c}$ intermediate
states is parametrized as
\begin{eqnarray}
\Lambda^\mu_{bs\gamma} &=& - \frac{G_{F}}{\sqrt{2}} \: V_{cb} V_{cs}^\ast \:
[ G_1^{c\overline{c}} (k^2) \: (k^2 g^{\mu\nu}- k^\mu k^\nu) \gamma_\nu L
\nonumber\\
& & + G_2^{c\overline{c}} (k^2) \: m_b i \sigma^{\mu\nu} k_\nu R ] .
\label{eq:10}
\end{eqnarray}
The interest here is in the $J/\psi$ contribution to the electromagnetic form
factors $G_{1,2}^{c\overline{c}}$. It will be derived from the weak vertex in
eq.\ \ref{eq:8} and the photon couplings shown in fig.\ 1. These couplings are
modeled after the $\gamma$-$\rho$ interaction, in the VMD description of the
electromagnetic properties of the nucleons \cite{VMD:Sakurai}. They correspond
to the gauge invariant interaction Lagrangian
\begin{equation}
{\cal L} = e Q_c \frac{f_{J/\psi}}{m_{J/\psi}} [ -\frac{1}{2} F_{\mu\nu}
\psi^{\mu\nu} - A_\mu (g^{\mu\nu} -
\frac{\partial^\mu \partial^\nu}{\Box}) J_\nu ] ,
\label{eq:11}
\end{equation}
where $A_\mu$ and $\psi_\mu$ are the photon and the $J/\psi$ fields;
$F_{\mu\nu} \equiv \partial_\mu A_\nu - \partial_\nu A_\mu$ and $\psi_{\mu\nu}
\equiv \partial_\mu \psi_\nu - \partial_\nu \psi_\mu$. The current $J_\nu$ is
that in eq.\ \ref{eq:9}. The second term on the RHS of eq.\ \ref{eq:11} is an
extension of the result in ref.\ \cite{VMD:Sakurai}. It encompasses the more
general case where the current is not necessarily conserved: in order to
preserve gauge invariance, only its conserved part was included in the
interaction. The phenomenological parameter $f_{J/\psi}$ is the same as that
defined before, since $J^\mu_{e.m.} = Q_c\: \overline{c}\gamma^\mu c + \cdots$.
In general, the two gauge invariant terms in the interaction Lagrangian would
have independent couplings. However, in the assumption of complete VMD
\cite{VMD:Sakurai}, the $k^2$ dependence of the electromagnetic form factors
is dominated by the vector meson pole, i.\ e.\ $G_{1,2}^{c\overline{c}}
\propto 1/(k^2 - m_{J/\psi}^2)$; this leads to the result in eq.\ \ref{eq:11}.
The $J/\psi$ contribution to the form factors $G_{1,2}^{c\overline{c}}$, in
the $bs\gamma$ vertex, is then
\begin{equation}
G_{1,2}^{J/\psi}(k^2) =  - \: (C_2 + \frac{1}{3} C_1) \: e Q_c
f_{J/\psi} m_{J/\psi} g_{1,2} \: \frac{1}{k^2 - m_{J/\psi}^2} ,
\label{eq:12}
\end{equation}
where $f_{J/\psi} \times g_{1,2}$ is taken to be constant in $k^2$, for
consistency with the complete VMD assumption. Corrections to this assumption,
due to the contribution of other $c$-$\overline{c}$ states (such as the open
charm continuum), are discussed later.

\begin{figure}[htbp]
\unitlength1mm
\begin{picture}(80,100)
\put(0,0){\makebox(80,80)
{\epsfig{figure=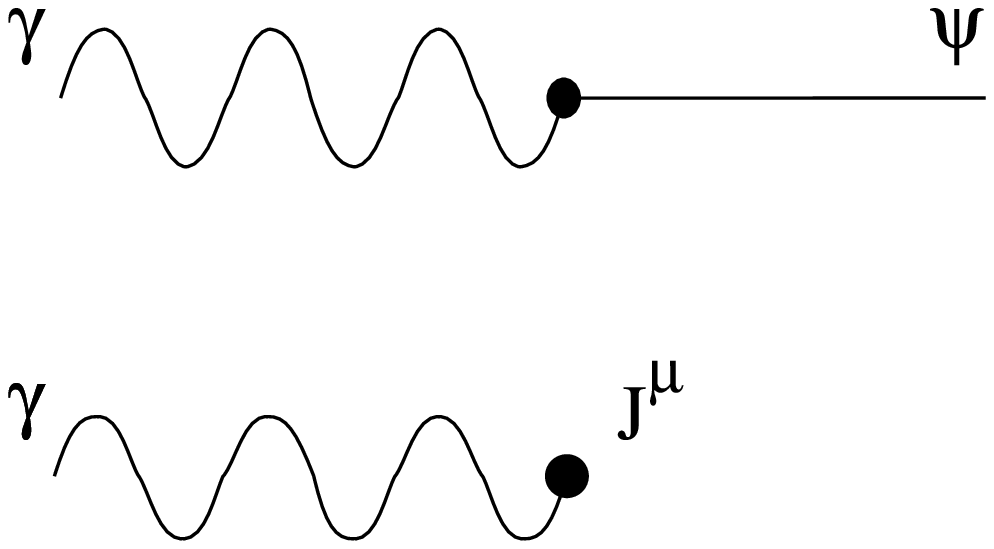,height=80mm,width=80mm}}}
\put(70,50){{\Large  $ - i e Q_c \: \frac{f_{J/\psi}}{m_{J/\psi}} \: k^2 \:
( g^{\mu\nu} - \frac{k^\mu k^\nu}{k^2} ) $}}
\put(70,10){{\Large $ - i e Q_c \: \frac{f_{J/\psi}}{m_{J/\psi}} \:
( g^{\mu\nu} - \frac{k^\mu k^\nu}{k^2} ) \: J_\nu $}}
\end{picture}
\caption{The photon vertices that correspond to the interaction Lagrangian
of eq.\ \protect\ref{eq:11}.}
\label{fig:1}
\end{figure}

\subsection{The $J/\psi$ contribution to the radiative decay amplitudes}

The amplitudes for the inclusive and exclusive radiative $B$ decays, due to
the $J/\psi$ contribution, follow from eqs.\ \ref{eq:10} and \ref{eq:12}. Only
the magnetic dipole moment type structure in the vertex (i.\ e.\ the form
factor $G_2^{J/\psi}$) contributes, when the photon is on-shell. For the
inclusive decay, the magnitude of the $J/\psi$ contribution is
\begin{eqnarray}
|A(b \stackrel{J/\psi}{\rightarrow} s \gamma)| &=&
| \frac{G_{F}}{\sqrt{2}} \: V_{cb} V_{cs}^\ast \: G_2^{J/\psi}(0) \: m_b
\: \varepsilon_{J/\psi}^{\mu\ast} <s| \overline{s} i \sigma_{\mu\nu} k^\nu R b
|b> | \nonumber \\
&=& | G_{F} \: V_{cb} V_{cs}^\ast \:
(C_2 + \frac{1}{3} C_1) \: 2 m_b^3 e Q_c \:
\frac{f_{J/\psi}}{m_{J/\psi}} \: g_2 | ;
\label{eq:13}
\end{eqnarray}
and for the exclusive $B \rightarrow K^\ast \gamma$ decay, it is
\begin{eqnarray}
|A(B \stackrel{J/\psi}{\rightarrow} K^\ast \gamma)| &=&
| \frac{G_{F}}{\sqrt{2}}
V_{cb} V_{cs}^\ast  \: G_2^{J/\psi}(0) \: m_b \:
\varepsilon_{J/\psi}^{\mu\ast} <K^\ast| \overline{s} i \sigma_{\mu\nu} k^\nu
R b |B> | \nonumber \\
&=& | \frac{G_{F}}{\sqrt{2}} \:  V_{cb} V_{cs}^\ast
(C_2 + \frac{1}{3} C_1)\: m_b e Q_c \:
\frac{f_{J/\psi}}{m_{J/\psi}} \: g_2 \nonumber\\
& & \times (m_B^2 - m_{K^\ast}^2) F_1(0) | .
\label{eq:14}
\end{eqnarray}
The form factor $F_1(k^2)$ is one of the three form factors that parametrize
the hadronic matrix element $<K^\ast|\overline{s} i \sigma^{\mu\nu} k_\nu R b
|B>$ in the decay amplitude. These form factors are defined in the Appendix.

The phenomenological model presented in here has the peculiarity that it gives
no contribution of the $J/\psi$ resonance to the radiative $B$ decays, when
the $b \rightarrow s J/\psi$ transition is treated within the factorization
approximation. In that approximation, as it was shown above, $g_2 = 0$, and so
there is no $J/\psi$ contribution to the magnetic dipole moment structure in
the $bs\gamma$ vertex. This result can be understood from a different
perspective. The Hamiltonian in eq.\ \ref{eq:5} gives a perturbative
contribution to $G_{1,2}^{c\overline{c}}$, that, at the lowest order,
corresponds to the $c$-quark loop diagram in fig.\ 2. This gives
$G_2^{c\overline{c}} = 0$ and $G_1^{c\overline{c}} = (C_2 + C_1/3) e Q_c
\Pi(k^2)$, where $\Pi(k^2)$ has a cut along the real axis starting at $k^2 =
4 m_c^2$, and no poles. This contribution is to be interpreted as an average
over the $c$-$\overline{c}$ resonant and continuum virtual states. In order to
obtain the poles, such as the $J/\psi$ pole, explicitly, soft QCD effects
would have to be included. In particular, the soft gluon exchanges between the
$c$-quark lines inside the loop would yield the $c$-$\overline{c}$ bound
states. This would result in including the $J/\psi$ pole in $\Pi(k^2)$, but
there would still be no contribution to $G_2^{c\overline{c}}$. The latter, and
the associated magnetic dipole moment structure of the $bs\gamma$ vertex, can
only appear due to gluon exchanges between the quark lines inside the loop and
the external quark lines, i.e. beyond the factorization approximation
\cite{g2:fact}.

\begin{figure}[htbp]
\centering
\unitlength1mm
\begin{picture}(100,100)
\put(0,0){\makebox(100,100)
{\epsfig{figure=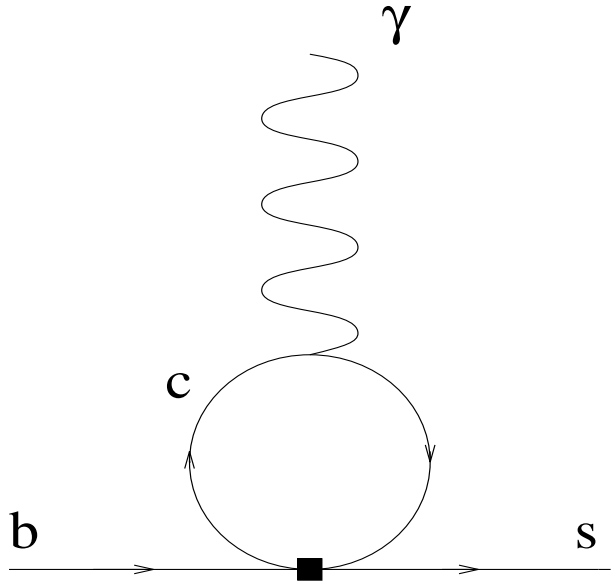,height=80mm,width=80mm}}}
\end{picture}
\caption{The lowest order perturbative contribution, from the
effective Hamiltonian in eq.\ \protect\ref{eq:5}, to the form factors
$G_{1,2}^{c\overline{c}}$ in the $bs\gamma$ vertex.}
\label{fig:2}
\end{figure}

\section{Quantitative estimates}

In order to obtain a quantitative estimate for the size of the $J/\psi$
contribution to the radiative $B$ decays, in eqs.\ \ref{eq:13} and \ref{eq:14},
it is necessary to determine the size of $g_2$, in the $bsJ/\psi$ vertex. One
possibility would be to extract $g_{1,2}(m_{J/\psi}^2)$ from the experimental
values for the branching ratio, $BR(B \rightarrow J/\psi + anything) = (1.15
\pm 0.07)\%$ \cite{Browder:Bdecays}, and the polarization, $\Gamma_L/\Gamma =
0.59 \pm 0.15$ \cite{Browder:Bdecays}, in the inclusive decay. This can only be
done after removing from the data the contribution from the $B$ decays into
$\psi^\prime$ and $\chi_{c1}$, that in turn decay into $J/\psi$. The effect on
the branching ratio has been measured, and $BR(b \rightarrow s J/\psi) = (0.82
\pm 0.08)\%$ \cite{Browder:Bdecays} for the direct decay; but the effect on
the polarization has not, and so $\Gamma_L/\Gamma$ for the direct decay could
range from 0.18 to 1. This large uncertainty is not the major obstacle in
extracting $g_{1,2}$ from the inclusive $b \rightarrow s J/\psi$ data; because
the longitudinal and transversal decay rates are not sensitive to the sign of
the corresponding amplitudes, $g_{1,2}$ can only be determined up to a 4-fold
ambiguity. The ambiguity is particularly serious for an estimate of
$g_2(m_{J/\psi}^2)$. For example, if the $J/\psi$-mesons from the cascade
decays are unpolarized, then $\Gamma_L/\Gamma = 0.69 \pm 0.21$ for the direct
decay; taking $|V_{cb}| = 0.038 \sqrt{1.63 {\rm psec}/\tau_b}$ \cite{CKM:Vcb}
and $m_b = 5.0$ GeV in
\begin{eqnarray}
\Gamma_{L,T}(b \rightarrow s J/\psi) &=&
\frac{1}{8\pi} G_F^2 |V_{cb}V_{cs}^\ast|^2
(C_2 + \frac{1}{3} C_1)^2 m_b (1 - \frac{m_{J/\psi}^2}{m_b^2})^2 \nonumber\\
& & \times \left\{ \begin{array}{ll}
[ g_1(m_{J/\psi}^2) - g_2(m_{J/\psi}^2) ]^2 m_b^2 m_{J/\psi}^2 & \mbox{(L)} \\
& \\
2 \: [ g_1(m_{J/\psi}^2) m_{J/\psi}^2 - g_2(m_{J/\psi}^2) m_b^2 ]^2
& \mbox{(T)}
\end{array}
\right.
\label{eq:15}
\end{eqnarray}
gives (up to a sign) $g_2(m_{J/\psi}^2) = 0.26 \pm 0.03$ or $0.04 \pm 0.07$,
which are very different in magnitude.

The alternative is to extract $g_{1,2}(m_{J/\psi}^2)$ from the data for the
exclusive decays. The branching ratio and the polarization for the $B
\rightarrow K^\ast J/\psi$ decay \cite{Browder:Bdecays},
\begin{eqnarray}
BR(B \rightarrow K^\ast J/\psi) &=& (1.64 \pm 0.27) \times 10^{-3}
\label{eq:16} \\
\left( \frac{\Gamma_L}{\Gamma} \right)_{B \rightarrow K^\ast J/\psi}
&=& 0.78 \pm 0.07 ,
\label{eq:17}
\end{eqnarray}
allow to determine $g_{1,2}$, up to the same 4-fold ambiguity as in the
inclusive case. But here the additional $B \rightarrow K J/\psi$ branching
ratio \cite{Browder:Bdecays},
\begin{eqnarray}
BR(B \rightarrow K J/\psi) &=& (0.089 \pm 0.013) \% , \label{eq:18}
\end{eqnarray}
can be used to reduce the ambiguity to that in the overall sign of $g_{1,2}$.
The longitudinal and transversal $B \rightarrow K^\ast J/\psi$ decay
amplitudes are
\begin{eqnarray}
\lefteqn{A_L(B \rightarrow K^\ast J/\psi) =
- \frac{G_F}{\sqrt{2}} V_{cb} V_{cs}^\ast (C_2 + \frac{1}{3} C_1)
\frac{m_{J/\psi}}{2 m_{K^\ast}}} \nonumber\\
& & \times \left\{ g_1(m_{J/\psi}^2) (m_B + m_{K^\ast})
\left[ A_1(m_{J/\psi}^2) (m_B^2 - m_{K^\ast}^2 - m_{J/\psi}^2) \right. \right.
\nonumber\\ & & \left. - A_2(m_{J/\psi}^2)
\frac{4 m_B^2 |\vec{k}|^2}{(m_B + m_{K^\ast})^2} \right] \nonumber\\
& & + g_2(m_{J/\psi}^2) m_b \left[ - F_2(m_{J/\psi}^2)
(m_B^2 + 3 m_{K^\ast}^2 - m_{J/\psi}^2) \right. \nonumber\\
& & + F_3(m_{J/\psi}^2) \left. \left.
\frac{4 m_B^2 |\vec{k}|^2}{m_B^2 - m_{K^\ast}^2} \right] \right\}
\label{eq:19}
\end{eqnarray}
and
\begin{eqnarray}
\lefteqn{A_T(B \rightarrow K^\ast J/\psi) =
- \frac{G_F}{\sqrt{2}} V_{cb} V_{cs}^\ast (C_2 + \frac{1}{3} C_1)
(m_B + m_{K^\ast})} \nonumber\\
& &  \times \sum_\pm \left\{ g_1(m_{J/\psi}^2) m_{J/\psi}^2
\left[ - A_1(m_{J/\psi}^2) \mp V(m_{J/\psi}^2)
\frac{2 m_B |\vec{k}|}{(m_B + m_{K^\ast})^2} \right] \right. \nonumber\\
& &  +  \left. g_2(m_{J/\psi}^2) m_b (m_B - m_{K^\ast})
\left[ F_2(m_{J/\psi}^2) \mp F_1(m_{J/\psi}^2)
\frac{m_B |\vec{k}|}{m_B^2 - m_{K^\ast}^2} \right] \right\} , \nonumber\\
& &
\label{eq:20}
\end{eqnarray}
respectively; the $B \rightarrow K J/\psi$ decay amplitude is
\begin{eqnarray}
A(B \rightarrow K J/\psi) &=&
- \frac{G_F}{\sqrt{2}} V_{cb} V_{cs}^\ast (C_2 + \frac{1}{3} C_1)
2 m_B |\vec{k}| m_{J/\psi} \nonumber\\
& & \times \left[ g_1(m_{J/\psi}^2) f_1(m_{J/\psi}^2) +
g_2(m_{J/\psi}^2) m_b s(m_{J/\psi}^2) \right]
\label{eq:21}
\end{eqnarray}
($|\vec{k}|$ is the  $J/\psi$ momentum in the $B$ rest-frame). The hadronic
matrix elements $<K^{(\ast)}| \overline{s} \gamma_\nu L b |B>$ and
$<K^{(\ast)}| \overline{s} i \sigma^{\mu\nu} k_\nu R b |B>$ have been
parametrized in terms of the form factors $V$, $A_{0,1,2}$, $f_{0,1}$ and
$F_{1,2,3}$, $s$, respectively, as defined in the Appendix. In order to
minimize the uncertainty that is inherent to any particular model for these
form factors, one can choose instead to relate them to the form factors that
can be measured in semileptonic decays. In ref.\ \cite{ffactors:IW}, Isgur and
Wise have used the Heavy Quark symmetry (HQS) to related the $B \rightarrow
K^{(\ast)}$ form factors to the form factors in $D \rightarrow K^{(\ast)} l
\overline{\nu}_l$; their method will be used in here, and the results are
summarized in the Appendix. It must be pointed out that these results are not
entirely model independent, as some assumption must be made regarding the
$k^2$ dependence of the form factors \cite{ffactors:BD}. The associated
uncertainty is hard to quantify and will not appear in the results, but it
should be kept in mind.

\begin{figure}[htbp]
\centering
\unitlength1mm
\begin{picture}(110,110)
\put(0,0){\makebox(110,110)
{\epsfig{figure=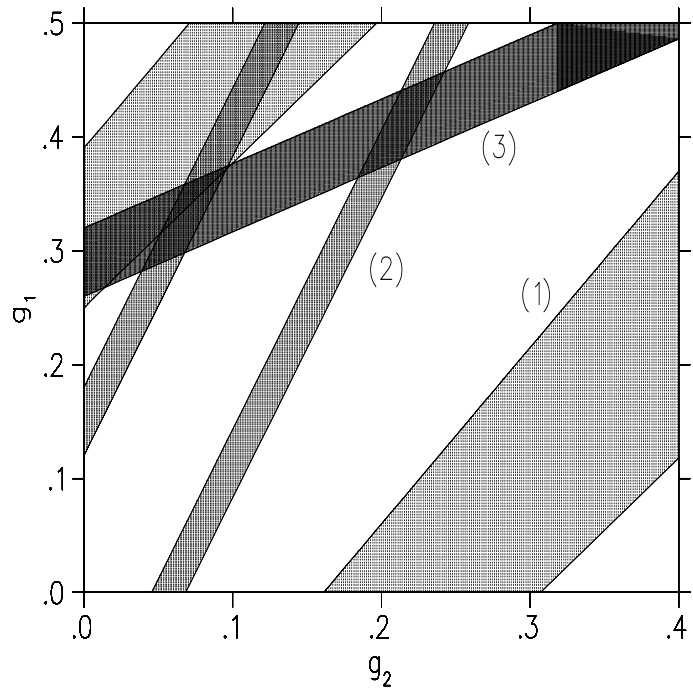,height=90mm,width=90mm,angle=90}}}
\end{picture}
\caption{Allowed region on the $(g_1,g_2)$ plane, from the data for the
longitudinal (1) and transversal (2) $B \rightarrow K^\ast J/\psi$ rates,
and for the $B \rightarrow K J/\psi$ rate (3).}
\label{fig:3}
\end{figure}

When compared to the experimental results, the magnitude of the amplitudes in
eqs.\ \ref{eq:19}--\ref{eq:21} give the straight lines
\begin{eqnarray}
g_1 = \pm a_i + b_i g_2 \hspace{.3in} (i = 1,2,3) ,
\label{eq:22}
\end{eqnarray}
in the $(g_1,g_2)$ plane (for the transverse amplitude the exact solution does
not give a straight line; but this is a very good approximation in the region
of interest). The parameters $a_i$ and $b_i$ are listed in table \ref{table:1};
the errors reflect the uncertainties in eqs.\ \ref{eq:16}--\ref{eq:18} and in
the normalization of the $B \rightarrow K^{(\ast)}$ form factors (see eq.\
\ref{eq:A16}, in the Appendix). The corresponding allowed regions are shown in
fig.\ \ref{fig:3}, and their overlap gives
\begin{equation}
|g_1(m_{J/\psi}^2)| = 0.31 {\rm -} 0.38 \hspace{.3in}
|g_2(m_{J/\psi}^2)| = 0.05 {\rm -} 0.10
\label{eq:23}
\end{equation}
(an ambiguity in the overall sign of $g_{1,2}$ remains). These results are
also sensitive to the values of the Wilson coefficients $C_{1,2}$ and of
$|V_{cb}| \sqrt{\tau_b}$ that were chosen. The associated errors, although
large, were not included as they will not affect the results that follow.

\vspace{.2in}
\begin{table}[htbp]
\centering
\begin{tabular}{ccc}
$i$ & $a_i$  &  $b_i$  \\ [.15in] \hline \\
1 & $0.32 \pm 0.07$ & $1.41 \pm 0.14$ \\
2 & $0.15 \pm 0.03$ & $2.63$ \\
3 & $0.29 \pm 0.03$ & $0.57$ \\ [.1in] \hline
\end{tabular}
\caption{The parameters for the lines $g_1 = \pm a_i + b_i g_2$
($i$ = 1,2,3) in fig.\ 3.}
\label{table:1}
\end{table}
\vspace{.2in}

Finally, the $J/\psi$ contribution to the radiative $B$ decay amplitudes, in
eqs.\ \ref{eq:13} and \ref{eq:14}, can be compared to the experimental values
for the full amplitudes, from eqs.\ \ref{eq:1} and \ref{eq:2}. For the
inclusive decay,
\begin{eqnarray}
\frac{|A(b \stackrel{J/\psi}{\rightarrow} s \gamma)|}
{|A(b \rightarrow s \gamma)|_{{\rm exp.}}} &=&
0.15 \pm 0.05 ;
\label{eq:24}
\end{eqnarray}
and, for the exclusive decay,
\begin{eqnarray}
\frac{|A(B \stackrel{J/\psi}{\rightarrow} K^\ast \gamma)|}
{|A(B \rightarrow K^\ast \gamma)|_{{\rm exp.}}} &=&
\frac{F_1(0)}{0.96} \: \times
(0.12 \pm 0.05) .
\label{eq:25}
\end{eqnarray}
As pointed out above, these results are not affected by the uncertainties in
$|V_{cb}| \sqrt{\tau_b}$ and in $|C_2 + C_1/3|$. The errors indicated
correspond to the uncertainties in eq.\ \ref{eq:23}, and in the experimental
branching ratios for the radiative decays. For the exclusive case, an
additional uncertainty is associated with the value of $F_1(0)$. In here,
$F_1(0) = 0.96 \pm 0.11$ (see Appendix), but it is smaller in other popular
models for the $B \rightarrow K^\ast$ form factors (in the BSW model
\cite{ffactors:BSW} $F_1(0) = 0.69$ and in the JW model \cite{ffactors:JW}
$F_1(0) = 0.59$). Finally, it should be pointed out that the sign of $g_2$
could not be determined; thus it cannot be said whether the long and short
distance contributions to the radiative $B$ decay amplitudes interfere
destructively or constructively.

\section{Conclusion}

A phenomenological model was constructed that describes the contribution to
the radiative $B$ decays from the tree level decay into the $J/\psi$ resonance,
followed by the $J/\psi$-photon conversion. To account for the weak decay, an
effective $bsJ/\psi$ vertex was introduced, which is used to describe both the
inclusive and the exclusive $B$ decays into $J/\psi$. This assumes that the
hadronization effects in the $J/\psi$ production and in the $B \rightarrow
K^{(\ast)}$ transition can be treated separately. The latter can then be
described in terms of the usual set of form factors, related to those in
semileptonic decays; whereas the former are described in terms of a new set of
form factors that are determined empirically. The $J/\psi$-photon transition
is modeled after the VMD ideas that were used, for example, to describe the
$\rho$-meson contribution to the nucleon electromagnetic form factors. The
assumption in here is that of complete $J/\psi$ dominance of the
electromagnetic form factors, i.\ e.\ the other $c$-$\overline{c}$
contributions are neglected. This leads to a Lagrangian for the
$J/\psi$-photon interaction, parametrized by the $J/\psi$ decay constant.

Within this model, the $J/\psi$ contribution to the $B$-meson radiative decays
was estimated to be $(10 {\rm -} 20)\%$ of the observed  $b \rightarrow s
\gamma$ amplitude, and $(7 {\rm -} 17)\%$  of the $B \rightarrow K^\ast
\gamma$ amplitude. The large uncertainties correspond mostly to experimental
errors and will be reduced in the future. There is however an additional
uncertainty from some degree of model dependence in extracting the form
factors in the $bsJ/\psi$ vertex from the data. Also not shown explicitly are
the errors inherent to the assumptions that underlie the phenomenological
model. In particular, the assumption of complete VMD is probably too strong.
It has been suggested \cite{Tsk:81} that the effect of $c$-$\overline{c}$
contributions other than the $J/\psi$-meson can be included in the formalism
derived from the complete VMD assumption, by allowing for an effective $k^2$
dependence of $f_{J/\psi}$. Within this prescription, the data for $J/\psi$
photoproduction and for charmonium radiative decays reveal a significant
departure from complete VMD \cite{fpsi:81} \cite{Desh:94}. In ref.\
\cite{fpsi:81}, it is found that
\begin{equation}
\frac{f_{J/\psi}(0)}{f_{J/\psi}(m_{J/\psi}^2)} \sim 0.6 ,
\label{eq:26}
\end{equation}
which should be viewed as a suppression factor that multiplies the results
given here. However, the size of this suppression is uncertain, and the $k^2$
dependence of the form factors in the $bsJ/\psi$ vertex remains unknown. For
these reasons, the results consistent with the complete VMD assumption were
reported in here, while corrections to this assumption await further work.

\section*{}

This work was partly supported by the Natural Science and Engineering
Research Council of Canada.

\section*{Appendix}

The hadronic matrix elements in the decay amplitudes are parametrized as
follows:
\begin{eqnarray}
\lefteqn{<K^\ast(p^\prime,\varepsilon^\prime)|\overline{s}\gamma^\mu L b|B(p)>
 = \frac{-1}{m_B + m_{K^\ast}} 2 i \epsilon^{\mu\alpha\beta\gamma}
\varepsilon_\alpha^{\prime\ast} p_\beta^\prime p_\gamma V(k^2)} \nonumber\\
& & -(m_B + m_{K^\ast}) \varepsilon^{\prime\mu\ast} A_1(k^2)
+\frac{\varepsilon^{\prime\ast}.k}{m_B + m_{K^\ast}}
(p+p^\prime)^\mu A_2(k^2)
\nonumber\\
& & + 2 m_{K^\ast} \frac{\varepsilon^{\prime\ast}.k}{k^2} k^\mu
[ A_3(k^2) - A_0(k^2) ],
\label{eq:A1}
\end{eqnarray}
where
\begin{equation}
2 m_{K^\ast} A_3(k^2) \equiv  (m_B + m_{K^\ast}) A_1(k^2)
- (m_B - m_{K^\ast}) A_2(k^2)
\label{eq:A2}
\end{equation}
and $A_0(0) =  A_3(0)$;
\begin{eqnarray}
\lefteqn{<K^\ast(p^\prime,\varepsilon^\prime)|\overline{s} i
\sigma^{\mu\nu} k_\nu R b|B(p)> = i \epsilon^{\mu\alpha\beta\gamma}
\varepsilon_\alpha^{\prime\ast} p_\beta^\prime p_\gamma F_1(k^2)}
\nonumber \\ & & +[(m_B^2 - m^2_{K^\ast}) \varepsilon^{\prime\mu\ast}
- \varepsilon^{\prime\ast}.k (p+p^\prime)^\mu] F_2(k^2) \nonumber \\
& & + \varepsilon^{\prime\ast}.k [k^\mu - \frac{k^2}{m_B^2 - m^2_{K^\ast}}
(p+p^\prime)^\mu] F_3(k^2) ,
\label{eq:A3}
\end{eqnarray}
where $F_1(0) = 2 F_2(0)$;
\begin{eqnarray}
<K(p^\prime)|\overline{s}\gamma^\mu L b|B(p)>
&=& (p + p^\prime)^\mu f_1(k^2) \nonumber\\
& & + \frac{m_B^2 - m_K^2}{k^2} k^\mu [ f_0(k^2) - f_1(k^2) ] ,
\label{eq:A4}
\end{eqnarray}
where $f_1(0) = f_0(0)$; and
\begin{eqnarray}
<K(p^\prime)| \overline{s} i \sigma^{\mu\nu} k_\nu R b |B(p)> &=&
s(k^2) [ (p + p^\prime)^\mu k^2 - (m_B^2 - m_K^2) k^\mu ]
\label{eq:A5}
\end{eqnarray}
($k = p-p^\prime$; $L,R \equiv 1\mp\gamma_5$).

In ref.\ \cite{ffactors:IW}, Isgur and Wise pointed out that in the static
b-quark limit $\gamma_0 b = b$, in the $B$-meson rest-frame, and so the
$<K^{(\ast)}| \overline{s}\gamma^\mu L b |B>$ and $<K^{(\ast)}| \overline{s}
i \sigma^{\mu\nu} k_\nu R b |B>$ form factors are related by
\begin{eqnarray}
F_1(k^2) &=& 2 (m_B - E_{K^\ast}) \frac{V(k^2)}{m_B + m_{K^\ast}}
+ \frac{m_B + m_{K^\ast}}{m_B} A_1(k^2) ,
\label{eq:A6}
\\
F_2(k^2) &=& \frac{2 m_B |\vec{p}_{K^\ast}|^2}{m_B^2 - m^2_{K^\ast}}
\frac{V(k^2)}{m_B + m_{K^\ast}} + \frac{m_B - E_{K^\ast}}{m_B - m_{K^\ast}}
A_1(k^2) ,
\label{eq:A7}
\\
F_3(k^2) &=& (m_B + E_{K^\ast}) \frac{V(k^2)}{m_B + m_{K^\ast}}
- \frac{m_B^2 - m^2_{K^\ast}}{m_B}
\{ \frac{V(k^2)}{m_B + m_{K^\ast}} \nonumber \\
& & + \frac{1}{2} \frac{1}{m_B - m_{K^\ast}} A_1(k^2)
- \frac{1}{2} \frac{1}{m_B + m_{K^\ast}} A_2(k^2) \nonumber\\
& & + \frac{m_{K^\ast}}{k^2}
[ A_3(k^2) - A_0(k^2) ] \}
\label{eq:A8}
\end{eqnarray}
(where $E_{K^\ast}$ and $|\vec{p}_{K^\ast}|$ are the energy and momentum
of the $K^\ast$ meson in the $B$ rest-frame), and
\begin{eqnarray}
s(k^2) &=& \frac{1}{2 m_B} \{ - f_1(k^2) +
\frac{m_B^2 - m_K^2}{k^2} [ f_0(k^2) - f_1(k^2) ] \} .
\label{eq:A9}
\end{eqnarray}
The $B \rightarrow K^{(\ast)}$ form factors $V$, $A_{0,1,2}$ and $f_{+,-}$ are
then related to the analogous $D \rightarrow K^{(\ast)}$ form factors, through
the HQS relations \cite{ffactors:IW}
\begin{eqnarray}
V(t^\ast) &=& \left( \frac{\alpha_s(m_b)}{\alpha_s(m_c)} \right)^{-6/25}
\sqrt{\frac{m_c}{m_b}} \: \frac{m_B + m_{K^\ast}}{m_D + m_{K^\ast}} \:
V^{DK^\ast}(0) ,
\label{eq:A10} \\
A_1(t^\ast) &=&  \left( \frac{\alpha_s(m_b)}{\alpha_s(m_c)} \right)^{-6/25}
\sqrt{\frac{m_b}{m_c}} \: \frac{m_D + m_{K^\ast}}{m_B + m_{K^\ast}} \:
A_1^{DK^\ast}(0) ,
\label{eq:A11} \\
A_2(t^\ast) &=& \frac{1}{2} \left( \frac{\alpha_s(m_b)}{\alpha_s(m_c)}
\right)^{-6/25} \sqrt{\frac{m_c}{m_b}}\:
\frac{m_B + m_{K^\ast}}{m_D + m_{K^\ast}} \:
\{ (1 + \frac{m_c}{m_b}) \:  A_2^{DK^\ast}(0)
\nonumber\\
& & + (1 - \frac{m_c}{m_b}) (m_D + m_{K^\ast}) 2 m_{K^\ast}
\left[ \frac{A_0^{DK^\ast}(k^2) - A_3^{DK^\ast}(k^2)}{k^2}  \right]_{k^2=0} \}
\nonumber\\
& & \label{eq:A12}
\end{eqnarray}
and
\begin{eqnarray}
f_1(t) &=& \frac{1}{2} \left( \frac{\alpha_s(m_b)}{\alpha_s(m_c)}
\right)^{-6/25} \sqrt{\frac{m_b}{m_c}} \: \{ (1 + \frac{m_c}{m_b}) f_1^{DK}(0)
\nonumber\\
& &  - (1 - \frac{m_c}{m_b}) (m_D^2 - m_K^2)
\left[ \frac{f_0^{DK}(k^2) - f_1^{DK}(k^2)}{k^2} \right]_{k^2=0} \} ,
\label{eq:A13}
\end{eqnarray}
with
\begin{eqnarray}
t^{(\ast)} = m_b^2 + m_{K^{(\ast)}}^2 -
\frac{m_b}{m_c}  \: (m_c^2 + m_{K^{(\ast)}}^2) .
\label{eq:A14}
\end{eqnarray}
The $D \rightarrow K^{(\ast)}$ form factors, at $k^2 = 0$, are extracted from
the $D \rightarrow K^{(\ast)} l \overline{\nu}_l$ data, assuming a monopole
$k^2$ dependence as in the BSW model \cite{ffactors:BSW} (see table
\ref{table:2} for the pole masses). They are \cite{ffactors:DK}
\begin{eqnarray}
V^{DK^\ast}(0) &=& 1.12 \pm 0.16  \hspace{.3in}
A_1^{DK^\ast}(0) = 0.61 \pm 0.05  \nonumber\\
A_2^{DK^\ast}(0) &=& 0.45 \pm 0.09  \hspace{.3in}
f_1^{DK}(0) = 0.77 \pm 0.04 .
\label{eq:A15}
\end{eqnarray}
For the other parameters, the values used in here are $m_b = 5.0$ GeV, $m_c =
(1.5 \pm 0.2)$ GeV, and $\Lambda_{\overline{MS}}^{(4)} = (250 \pm 50)$ MeV
\cite{Pdg:94}.

The $k^2$ dependence of the  $B \rightarrow K^{(\ast)}$ form factors is not
determined by the HQS relations. As for the $D \rightarrow K^{(\ast)}$ form
factors, it will be assumed that it is the monopole dependence of the BSW
model (see also refs.\ \cite{ffactors:GKP} and \cite{ffactors:CDDGFN}). The
pole masses are given in table \ref{table:2}, and
\begin{eqnarray}
V(0) &=& 0.73 \pm 0.13  \hspace{.3in}
A_1(0) = 0.30 \pm 0.03  \nonumber\\
A_2(0) &=& 0.31 \pm 0.05  \hspace{.3in}
f_1(0) = 0.50 \pm 0.03 ,
\label{eq:A16}
\end{eqnarray}
from eqs.\ \ref{eq:A10}--\ref{eq:A13}.

\vspace{.3in}
\begin{table}[htbp]
\centering
\begin{tabular}{ccccccc}
& $V$  & $A_{1,2}$ & $A_0$ & $f_1$ & $f_0$ \\ [.15in] \hline \\
$D \rightarrow K^{(\ast)}:$
& 2.11 & 2.53  & 1.97  & 2.11  & 2.60  \\
$B \rightarrow K^{(\ast)}:$
& 5.43 & 5.82  & 5.38  & 5.43  & 5.89  \\ [.1in] \hline
\end{tabular}
\caption{The pole masses \protect\cite{ffactors:BSW} for the $B,D
\rightarrow K^{(\ast)}$ form factors.}
\label{table:2}
\end{table}


\pagebreak
\pagestyle{empty}
\section*{Figure Captions}
\begin{tabbing}
\=Figure 1: \= The photon vertices that correspond to the interaction
Lagrangian \\
\>	\>of eq.\ \ref{eq:11}.\\
\>      \> \\
\>Figure 2: \> The lowest order perturbative contribution, from the
effective\\
\>      \>Hamiltonian in eq.\ \ref{eq:5}, to the form factors
$G_{1,2}^{c\overline{c}}$ in the $bs\gamma$ vertex.\\
\>      \> \\
\>Figure 3: \> Allowed region on the $(g_1,g_2)$ plane, from the data for the
longitudinal (1)\\
\>      \>and transversal (2) $B \rightarrow K^\ast J/\psi$ rates, and for
the $B \rightarrow K J/\psi$ rate (3).\\
\>      \>
\end{tabbing}

\vspace{.8in}

\section*{Table Captions}
\begin{tabbing}
\=Table 1: \= The parameters for the lines $g_1 = \pm a_i + b_i g_2$
($i$ = 1,2,3) in fig.\ 3.\\
\>      \> \\
\>Table 2: \> The pole masses \cite{ffactors:BSW} for the $B,D \rightarrow
K^{(\ast)}$ form factors.\\
\>      \>
\end{tabbing}

\pagebreak
\pagestyle{empty}
\section*{ }
\begin{center}
Table 1\\
\vspace{.2in}
\begin{tabular}{ccc}
$i$ & $a_i$  &  $b_i$  \\ [.15in] \hline \\
1 & $0.32 \pm 0.07$ & $1.41 \pm 0.14$ \\
2 & $0.15 \pm 0.03$ & $2.63$ \\
3 & $0.29 \pm 0.03$ & $0.57$ \\ [.1in] \hline
\end{tabular}
\end{center}

\vspace{1in}

\begin{center}
Table 2\\
\vspace{.2in}
\begin{tabular}{ccccccc}
& $V$  & $A_{1,2}$ & $A_0$ & $f_1$ & $f_0$ \\ [.15in] \hline \\
$D \rightarrow K^{(\ast)}:$
& 2.11 & 2.53  & 1.97  & 2.11  & 2.60  \\
$B \rightarrow K^{(\ast)}:$
& 5.43 & 5.82  & 5.38  & 5.43  & 5.89  \\ [.1in] \hline
\end{tabular}
\end{center}


\vspace{.2in}


\begin{thebibliography}{Abcd:efgh}

\bibitem{CLEO:exclusive}
R. Ammar, {\em et al.} (CLEO Collaboration), Phys. Rev. Lett. {\bf 71}, 674
(1993).

\bibitem{CLEO:inclusive}
M. S. Alam, {\em et al.} (CLEO Collaboration), Phys. Rev. Lett. {\bf 74}, 2885
(1995).

\bibitem{qcd:bsgamma}
M. A. Shifman, A. I. Vainshtein, and V. I. Zakharov, Phys. Rev. D {\bf 18},
2583 (1978); S. Bertolini, F. Borzumati and A. Masiero, Phys. Rev. Lett.
{\bf 59}, 180 (1987); N. G. Deshpande {\it et al.}, Phys. Rev. Lett. {\bf 59},
183 (1987).

\bibitem{lla:bsgamma}
For a review of the calculation of the short distance contribution, with the
QCD corrections included in the leading logarithm approximation, see
M. Ciuchini {\em et al.}, Nucl. Phys. {\bf B421}, 41 (1994) or R. Grigjanis
{\em et al.}, Phys. Rep. {\bf 228}, 93 (1993), and references therein.

\bibitem{nlo:bsgamma}
A. J. Buras {\em et al.}, Nucl. Phys. {\bf B424}, 374 (1994).

\bibitem{GP:88}
E. Golowich and S. Pakvasa, Phys. Lett. {\bf B205}, 393 (1988).

\bibitem{GP:94}
E. Golowich and S. Pakvasa, Phys. Rev. {\bf D 51}, 1215 (1995).

\bibitem{Cheng:94}
H.-Y. Cheng, report no. hep-ph 9411330.

\bibitem{DTP:88}
N. G. Deshpande, J. Trampetic and K. Panose, Phys. Lett.
{\bf B214}, 467 (1988).

\bibitem{Desh:94}
N. G. Deshpande, X.-G. He and J. Trampetic, report no. hep-ph 9412222.

\bibitem{VMD:Sakurai}
See, for example, J. J. Sakurai, in {\em Currents and Mesons} (The Univ. of
Chicago Press, Chicago, 1969) chap. III.

\bibitem{Other:94}
Other than the works already indicated, see also
M. Ahmady, D. Liu and Z. Tao, report no. hep-ph 9302209;
D. Atwood, B. Blok and A. Soni, report no. hep-ph 9408373.

\bibitem{Pdg:94}
Particle Data Group, Phys. Rev. {\bf D 50}, 1173 (1994).

\bibitem{SD:QCD}
F. J. Gilman and M. B. Wise, Phys. Rev. {\bf D 20}, 2392 (1979).

\bibitem{fact:bspsi}
For a recent review of the factorization procedure in the
hadronic weak decays of $B$-mesons, see, for example,
M. Neubert {\em et al.}, in {\em Heavy Flavors}, ed. by A. J. Buras and
L. Lindner (World Scientific, Singapore, 1992).

\bibitem{g2:fact}
In principle, gluon exchanges between the external $b$ and $s$ quark lines
can be accomodated within the factorization approximation. They give
corrections to the $bs\gamma$ vertex that include terms of the magnetic dipole
moment type. However, such contributions to $G_2^{c\overline{c}}$ are still
multiplied by the factor $(k^2 g^{\mu\nu}- k^\mu k^\nu)$ from the lowest order
vertex, and so they will vanish for an on-shell photon.

\bibitem{Browder:Bdecays}
T. E. Browder and K. Honscheid, report no. hep-ph 9503414, and references
therein.

\bibitem{CKM:Vcb}
A. Ali and D. London, Z. Phys. {\bf C65}, 431 (1995).

\bibitem{ffactors:IW}
N. Isgur and M. B. Wise, Phys. Rev. {\bf D 42}, 2388 (1990).

\bibitem{ffactors:BD}
This may be improved in the future, when the form factors in $B \rightarrow
\pi (\rho) l \overline{\nu}_l$ can be measured, and related to those in here
through the flavor-$SU(3)$ symmetry; see G. Burdman and J. Donoghue, Phys.
Lett. {\bf B270}, 55 (1991).

\bibitem{ffactors:BSW}
M. Wirbel, B. Stech and M. Bauer, Z. Phys. {\bf C29}, 637 (1985);
{\bf C34}, 103 (1987).

\bibitem{ffactors:JW}
W. Jaus, Phys. Rev. {\bf D 41}, 3394 (1990); W. Jaus and D. Wyler,
Phys. Rev. {\bf D 41}, 3405 (1990).

\bibitem{Tsk:81}
K. Terasaki, Nuov. Cim. {\bf 60 A}, 475 (1981).

\bibitem{fpsi:81}
E. Paul, in {\em Proc. 1981 Int. Symp. on Lepton and Photon Interactions at
High Energies}, ed. W. Pfiel (University of Bonn, Bonn, 1981), p. 301.

\bibitem{ffactors:DK}
M. S. Witherell, invited talk at the Int. Symp. on Lepton and Photon
Interactions at High Energies, Ithaca, N.Y., 1993.

\bibitem{ffactors:GKP}
M. Gourdin, A. N. Kamal and X. Y. Pham, report no. hep-ph 9405318.

\bibitem{ffactors:CDDGFN}
R. Casalbuoni {\em et al.}, Phys. Lett. {\bf B299}, 139 (1993);
A. Deandrea {\em et al.}, Phys. Lett. {\bf B318}, 549 (1993).


\end{thebibliography}
\end{document}